\documentclass[showpacs,aps,pre,superscriptaddress,floatfix]{revtex4}
\usepackage{graphicx} 
\usepackage{amssymb}
\begin{document}  

\title{Non-universal non-equilibrium critical dynamics with disorder}

\author{M. D. Grynberg}
\author {G. L. Rossini} 

\affiliation{Departamento de F\'{\i}sica, Universidad Nacional
de La Plata, (1900) La Plata, Argentina}

\author{R. B. Stinchcombe}

\affiliation{Rudolf Peierls Centre for Theoretical  Physics, 
University of Oxford,1 Keble Road, Oxford OX1 3NP, UK}

\begin{abstract}
We investigate finite size scaling aspects of disorder reaction-diffusion
processes in one dimension utilizing both numerical and analytical 
approaches. The former averages the spectrum gap of the associated
evolution operators by doubling their degrees of freedom, while the
latter uses various techniques to map the equations of motion to a first 
passage time process. Both approaches are consistent with nonuniversal 
dynamic exponents, and with stretched exponential scaling forms 
for particular disorder realizations.
\end{abstract}


\maketitle

\section{Introduction} 

As the concept of dynamic ensemble distribution remains elusive, 
a way forward to the study of nonequilibrium systems relies partly on 
prototypical stochastic models [\onlinecite{Odor}]. The identification 
of scaling regimes for these latter at both large time and length scales 
has permitted to characterize a vast amount of nonequilibrium processes
in terms of universality classes [\onlinecite{Odor}].  However, their 
asymptotic dynamics can change completely in the presence of quenched
disorder or under space-dependent external forces [\onlinecite{Igloi}]. 
This is because particle motion may well remain localized around  strong 
spatial heterogeneities which prevail over stochastic fluctuations 
[\onlinecite{Igloi,Golosov}].

In this context, as well as for their value in modelling real non-equilibrium 
behaviour, reaction-diffusion models have been much studied in recent 
years, with such methods as real space renormalization,  yielding very 
detailed analytical results at large times [\onlinecite{Le-Doussal}]. In the 
thermodynamic limit, some universal and non-universal types of behavior
have been reported for these systems [\onlinecite{Gunter}] and related
models [\onlinecite{Krapivsky}], yet comparatively little is known on their 
relaxation times in finite disordered samples. In this work we focus on the 
dynamic exponents  associated to these time scales as well as on finite size 
scaling aspects which ultimately will turn out to be nonuniversal.  Although 
these processes are here schematized as one dimensional annihilating random
walks in a Brownian potential [\onlinecite{Le-Doussal,Gunter}], we expect 
them to be relevant in the description of such real cases as exciton dynamics
on long disordered polymers, given the experimental success of their 
homogeneous counterparts [\onlinecite{experiment}]. 

\begin{figure}[htbp]
\vskip -2cm
\centering
\includegraphics[width=0.6\textwidth]{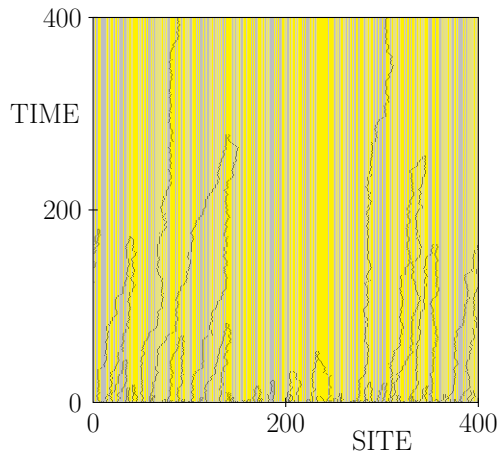}
\vskip -7cm
\caption{(Color online) Evolution snapshot, corresponding to a system with 
400 sites and periodic boundary conditions, with instantaneous reaction 
$R_l=1$ and a binary distribution of diffusion biases $h_l - h_l' \in 
\{0.6,\,-0.2\}$ with probability $p=0.5$ (regions in gray correspond to $h_l -
h_l' =0.6$).  It can be seen that particles tend to be trapped at boundaries
separating positive bias at the left from negative ones to the right.}
\label{snap}
\end{figure}

Typically, the microscopic dynamical rules of these models involve 
hard-core particles which hop randomly and annihilate in pairs on adjacent
locations.  To determine the effect of quenched disorder on dynamic 
exponents, here we allow for both varying diffusion and reaction rates in a
linear chain. So we consider the following one-step processes.  A particle at 
site $l\,$ $(l+1)\,$, chosen from $L$ locations, hops with rate $h_l\,$ $(h'_l)\,$
to site $l+1\,$ $(l\,)$ provided it is vacant. Independently, reacting attempts 
take place with rate $R_l$ whenever two particles occupy sites  $l$ and $l+1$.
As we shall see in Secs. III and IV, the nonuniversal situation alluded to above 
will be associated to random orientation changes in diffusion biases $h-h'$. 
This scenario is illustrated in Fig.\,\ref{snap}, where an evolution snapshot of 
these stochastic rules is depicted for the case of instantaneous  reactions 
under a binary distribution of biases.

Despite this simplicity, such modeling entails a variety of interesting
aspects which play a crucial role in the subsequent analysis. These are:
the nonequilibrium character of the process, formalisable in terms of a 
non-Hermitian quantum Hamiltonian [\onlinecite{Robin}]; 
criticality - associated with large (domain) scales,  and small energies; 
disorder - the distributions are the source of the remarkable non 
universality; particle nonconservation - bringing new features to the
non-universal critical phenomena through the interplay between particle
and hole sectors. 

This work is organized as follows. In Sec. II we present a calculation
approach that exploits the quantum analogy and focuses on the spectrum 
{\it gap} (real part) of the associated non-Hermitian Hamiltonian. For 
numerical convenience  the difficulties brought in by the particle 
nonconservation are remedied by duplicating the degrees of freedom of the
original process.   
Sec. III outlines an analytic treatment (cf. [\onlinecite{Dyson,Carlos}]), 
developed from equations of motion mappings combined with renormalisation
and transfer matrix techniques generalised to the non-conserving 
non-equilibrium nature of the system. Among the main features and 
procedures (briefly presented below) are: states appearing at quantised 
values of a complex phase obtained from phase steps for each bond and sign 
alternations at limiting phases; identification and aggregation of phase steps, 
as generalised random walks, in general biased, in the region between the 
limiting phases;  this makes the determination of phases and their 
distributions a first passage problem [\onlinecite{Redner}] in which both the 
character (power law or stretched exponential) of the non-universal critical 
dynamics and in particular the critical exponents are determined by the 
effective diffusion constant and bias, inheriting their parameter-dependences
(non-universality).
Sec. IV  implements numerically the ideas developed in Sec. II to check out the 
theoretical results obtained in Sec. III.  The computation of finite size scaling 
properties of the gap yield the dynamic exponents $z$, thus providing the 
fundamental scaling between length and time. In addition, numerical 
simulations are carried out to corroborate the nonuniversal findings for $z$.   
Sec. V contains a  summarizing discussion along with some remarks on 
extensions of this work.

\section{Fermion doubling}
The common starting point for both theoretical and numerical approaches
(Secs. III and IV respectively),  is the representation of the evolution operator 
of the stochastic process, described above, as a quantum spin-$1\over2$ non 
Hermitian Hamiltonian [\onlinecite{Robin}], which henceforth we recast as a
problem of  spinless fermions  via a Jordan-Wigner  transformation (JW) 
[\onlinecite{JW}]. Furthermore, these particles become actually {\it free} 
so long as the constraint  
\begin{equation}
\label{constraint}
R_l = h_l + h'_l\,,
\end{equation}
 is imposed for all lattice bonds [\onlinecite{Robin,GS}]. Specifically, it can be 
readily checked that in terms of fermion operators  $C_l, C^{\dag}_l$, the 
evolution operator under these conditions reads
\begin{equation}
\label{Hamiltonian}
H_C = \sum_{l=1}^L \, (h_l + h'_{l-1})\, C^{\dag}_l \,C_l
\,+ \,\sum_{l=1}^{L-1} \big( \, R_l \, C_l\,C_{l+1} 
- h_l \,C^{\dag}_{l+1} C_l -
h'_l \, C^{\dag}_l \,C_{l+1}  \,\big)\,+  B_C \,,
\end{equation}
($h'_0 \equiv h'_L$). According to the parity $e^{i \pi {\cal N}_C} \equiv 
e^{i \pi  \sum_j  C^{\dagger}_j C_j} $ of the subspace considered, here 
\begin{equation}
\label{boundary}
B_C = e^{i \pi {\cal N}_C}  \left(\,  h_L\, C^{\dag}_1\, C_L +  
h'_L C^{\dag}_L \,C_1 - R_L C_L\, C_1\, \right)\,,
\end{equation}
is a boundary term stemming from the JW  transformation under periodic 
boundary conditions (PBC).

The bilinear form imposed throughout by free fermion constraints is already
very suitable for theoretical analysis (as given in Sec. III), however the
pairing terms of $H_C$ would pose severe size restrictions in its numerical
analysis, demanding the consideration of matrices growing exponentially with 
the length of the system. In this respect, notice that a standard Bogoliubov 
transformation [\onlinecite{LSM}] is not effective within our disordered non 
Hermitian context. Rather, here we follow the strategy of 
Ref.\,[\onlinecite{MC}] in formally related systems and proceed to double the 
degrees of freedom of the original problem by  introducing a replica $H_D$ of 
the Hamiltonian $H_C$. The idea is to find  a simple (i.e. disorder 
independent) unitary transformation so as to cancel out all non-conserving 
particle terms of  the replicated system $H_{_{C+D}} = H_C + H_D$. To this 
end, we put forward the following fermion operators
\begin{eqnarray}
\nonumber
f_l &=& a_1^*\, C_l + a_2 \,C_l^{\dag} + b_1^* \,D_l + b_2 \,D_l^{\dag}\,,\\
\label{doubling}
g_l & =& a_3^* \,C_l + a_4 \,C_l^{\dag} + b_3^* \,D + b_4 \,D_l^{\dag}\,,
\end{eqnarray}
with $a$'s and $b$'s coefficients such that all pairing terms are ultimately
canceled in the $f-g$ representation. In fact, this can be thought of as a local 
Bogoliubov transformation in the real space of the double system. After some 
simple but lengthy algebra, we find that these coefficients should satisfy
\begin{eqnarray}
\nonumber
a_j^2 + b_j^2 &=& 0\,, \;\; 1 \le j \le 4\,,\\
\nonumber
a_1 \,a_3 + b_1 \,b_3 &=& 0\,,\;\; a_2 \,a_4 + b_2 \,b_4 = 0\,,\\
a_2^*\,a_1 + b_2^* \,b_1 &=& 0\,,\,\;\;a_2^*\,a_3 + b_2^*\,b_3 = 0\,,\\
\nonumber
a_1^*\, a_4 + b_1^*\,b_4 &=& 0\,,\;\; a_4^*\,a_3 + b_4^*\,b_3 = 0\,.
\end{eqnarray}
Among the various solutions we choose,
\begin{eqnarray}
\nonumber
a_1 &=& a_2 = a_3 = -\,a_4 = 1/2\,\\
b_1 &=& -\,b_2 = b_3 = b_4 = -\,i/2\,,
\end{eqnarray}
which leave us with a Hamiltonian of real parameters given by
\begin{eqnarray}
\label{doubleH}
H_{_{C+D}} &=& B_{_{C+D}} \,+ \,\sum_{l=1}^{L-1} \, \left [\,
h_l \left(\, f^\dag_l\,f_{l+1} - {\rm h.c.}   \,\right) + 
h'_l  \left(\, g^\dag_{l+1}\,g_l - {\rm h.c.}   \,\right) -
R_l  \left(  f^\dag_{l+1} \,g_l \,+ \, {\rm h.c.}  \, \right) \,\right] \\
\nonumber
&+& \sum_{l=1}^L \, \left[\, ( h_l + h'_{l-1} ) \, \left( \,1 + f_l^\dag\, g_l  
+ {\rm h.c.} \, \right)\,\right] \,.
\end{eqnarray}
As before, here $B_{_{C+D}}$ parallels the boundary terms referred
 to in Eq.\,(\ref{boundary}), which in this language take the form  
\begin{eqnarray}
\nonumber
B_{_{C+D}} &=&\frac{1}{2}  \left( e^{i \pi {\cal N}_C} +  
e^{i \pi {\cal N}_D} \right)  \left [\, h_L \, \left(   f^\dag_1 \,f_L - {\rm h.c.} \,
\right) + h'_L  \,\left(\, g^\dag_L\,g_1 - {\rm h.c.}   \,\right)
+ R_L \, \left(  f^\dag_1 \,g_L + {\rm h.c.}   \,\right) \,\right]\\
\label{dboundary}
&+& \frac{1}{2}  \left( e^{i \pi {\cal N}_C} -  e^{i \pi {\cal N}_D} \right) 
\left [\, h_L \, \left(   f^\dag_1 \,f^\dag_L - {\rm h.c.} \,
\right) + h'_L  \,\left(\, g^\dag_1\,g^\dag_L - {\rm h.c.}   \,\right)
+ R_L \, \left(  g_L^\dag\, f^\dag_1  + {\rm h.c.}   \,\right) \,\right]\,.
\end{eqnarray}
Therefore, the number of  $f$ and $g$ particles is {\it conserved}
as all pairing terms actually cancel out, provided 
we restrict attention to 
subspaces having $(-1)^{{\cal N}_C} = (-1)^{{\cal N}_D}$.

In passing, it is instructive to check how the exact solution of the 
homogeneous situation  is recovered by Fourier transforming the $f_l$ 
and $g_l$ operators respectively to a set of new  wave fermions  
$f_q,\, g_q$.  To account for the parity conserving terms implicated in 
$B_{_{C+D}}$, the $q$-momenta must be those of the set 
$Q^+ = \left\{ \pm \pi/L,\, \pm 3 \pi/L,\, \cdots\,,\pm (L-1) \pi/L \right\}$ 
for the even  subspace, while for the odd one the $q$'s should belong to  
$Q^- = \left\{ \pm 2 \pi/L,\, \pm 4 \pi/L,\, \cdots\,,\pm (L-2) \pi/L, 0,
\pi \right\}$,  say for $L$ even. After straightforward algebraic steps, it is 
readily found that the ordered Hamiltonian $H_o$  [\,i.e. $R_l = R, \,h_l = h,$ 
and $h'_l = h'\,\; \forall\, l\,$ in Eq.\,(\ref{doubleH})\,] reduces to 
\begin{equation}
H_o = R\, L + \sum_q \left(\begin{array}{ccc} 
 f_q^\dag  & , &g_q^\dag \end{array}\right)
\left[\begin{array}{cc} i \,2 h \sin q & R \,(1 - e^{-iq} ) \\ \\
R \,(1 - e^{iq} ) & - \,i \,2 h' \sin q  \end{array}\right]\,
\left(\begin{array}{c} f_q \\ \\ g_q \end{array}\right)\,.
\end{equation}
Using a simple non unitary $2 \times 2$ similarity transformation to new 
$F_q^{\pm}, G_q^{\pm}$ fermion like operators [\onlinecite{algebra}], 
the double system is finally diagonalized as
\begin{eqnarray}
\nonumber
H_o &=& R\, L + \sum_{q \in Q^{\pm} }\Lambda_q \,F^+_q \,F^-_q \,-
\, \sum_{q \in Q^\pm} \Lambda_{-q}\, G^+_q \, G^-_q\,,\\
\Lambda_q &=& R\, (1 - \cos q) + i\, (h-h') \sin q\,.
\end{eqnarray}
Thus, in doubling  the degrees of freedom  the $\Lambda_q$ excitations of 
the original problem are restored [\onlinecite{GS}], but now they appear 
forked in two symmetrical branches $\Lambda_q, -\Lambda_{-q}$ around a 
Fock  vacuum with "energy" $R\,L$  rather than with vanishing value, as it
would  correspond to the stationary state of the original stochastic $H_C$. 
Hence, in general the doubling transformation (\ref{doubling}) should not be 
regarded as a similarity one, although it preserves all the anticommutation 
relations which is enough to maintain the spacing of eigenlevels, and in 
particular the spectrum gap. In the homogeneous case, this scales
as $1/L^2$, so the usual dynamic exponent $z=2$ is restituted.

For the more interesting disordered situation, if we think of  $\{f_1, g_1, \, 
\cdots\,,  f_L, g_L\}$  as the $2 L$ single particle states of the replicated 
Hamiltonian  $H_{_{C+D}}$, then the  single excitations of 
Eq.\,(\ref{doubleH}) can be obtained by diagonalizing the real non 
symmetric five-diagonal matrix (with boundary conditions)

\begin{equation}
\nonumber
M_{\pm} = \left[\begin{array}{cccccccc}
{\cal R} & h_1 + h'_L & h_1 &0 &   \cdots & 0 & 
\pm  h_L &  \pm  R_L \\  \\
h_1 + h'_L  & {\cal R} & -R_1& -h'_1 &0 & \cdots &  0 
&  \mp h'_L \\ \\
-h_1 & -R_1 & {\cal R} & h_2 + h'_1 & h_2 & 0 & \cdots & 0 \\
0 & h'_1 & h_2 + h'_1 & {\cal R}& -R_2 & -h'_2 &0 & \vdots   \\
& \vdots &   &  & \hskip -1.5cm  \ddots &    & \vdots &   \\
\vdots  &0 & -h_{L-2} & -R_{L-2} & {\cal R}  & h_{L-1} + h'_{L-2} & 
h_{L-1} &0 \\ \\
0 & \cdots & 0 & h'_{L-2} & h_{L-1} + h'_{L-2} &{\cal R} 
& -R_{L-1} &-h'_{L-1}\\  \\
 \mp h_L & 0& \cdots &0 &-h_{L-1}& -R_{L-1}  
& {\cal R} & h_L + h'_{L-1}  \\ \\
 \pm R_L &  \pm  h'_L &  0&\cdots & 0 & h'_{L-1} 
& h_L + h'_{L-1} & {\cal R}
\end{array}\right]\,,
\label{matrices}
\end{equation}
where ${\cal R} = \sum_j R_j$. The numerical diagonalization of $M_{\pm}$ 
[\onlinecite{NR}] reveals the emergence, as before, of two symmetrical 
branches  $\{{\cal R} + \Lambda\},\,\{{\cal R} - \Lambda\}$ of single 
excitations around a ground state energy ${\cal R}$. Denoting by 
$\Lambda^{\pm}_0$ the lowest level excitation in either subspace 
(\,${\rm Re}\; \Lambda_0^{\pm} = 
{\rm min \; \{ Re}\; \Lambda^{\pm}\}\,)$, the spectrum gap $g$ of the 
original evolution operator (\ref{Hamiltonian}) with an initially even (odd) 
number of reacting-diffusing particles, is finally obtained as 
$g = 2\, {\rm Re}\; \Lambda^+_0,$ with a doubly degenerate
$\Lambda^+_0$  ($g = {\rm Re}\; \Lambda^-_0,\;$ with $\Lambda^-_0$ 
being non degenerate). 

We will recapitulate these ideas in the numerical diagonalizations
of Sec. IV so as to account for the theoretical findings of the next section.

\section{Phase steps as random walks}

In what follows we reconsider the 
non-conserving fermion Hamiltonian (\ref{Hamiltonian}), but without 
fermion doubling. One of the eigenvalue equations associated to that 
operator involves only "hole" state amplitudes $a_l$ at each site $l$. 
These provide an inhomogeneous term in the other equation, for the  
"particle" states $a^{+}_l$.  Using two-component vectors, the 
eigenvalue equation involves $2 \times 2$ matrix coefficients which 
depend on the excitation "energy"  $\Lambda$, and on bond-dependent 
biased hopping and bond-independent pair annihilation rates $h_l, 
h'_l, R_l$, constrained as in Eq.\,(\ref{constraint}), to keep the free-fermion 
character. The product of the associated $4 \times 4$ transfer matrices, 
$\prod_{l=1}^{L} T_l $, provides the spectrum through its trace, 
or appropriate matrix elements (depending on boundary conditions), 
effectively corresponding to quantisation of an aggregated phase.
One $2 \times 2$ part of $T_l $ is the transfer matrix $t_l$ for the 
self-contained subspace ("hole sector") provided by the states $a_l$.

From the coupled eigenvalue equations one may equivalently proceed
via the corresponding $4 \times 4$ transfer matrix $T_l$, or via 
renormalisation using a decimation method, or by mapping ratios of 
successive amplitudes. Here we mainly employ the last approach. The transfer 
matrix and decimation approaches are very instructive, in clarifying 
respectively the role of accumulating complex phases (which provide the
spectrum) and the renormalization of quenched probability distributions
of such aggregated variables, but in what follows it is sufficient and convenient 
to begin with the transfer matrix and then move over to the mapping approach.

For orientation and to obtain certain characteristic parameters 
it is useful to separate the rates (and consequently the transfer matrix
itself) into reference uniform, $l$-independent, "effective medium" 
parts ($h, h'=R-h$, indicated by the absence of a site label), together 
with the further random part involving a disorder strength $\delta$, 
and (binary) random variables $\zeta_l$ at each bond.  
Then $T_l$ becomes  $T_l \equiv T + \hat{\Delta}_l$.
For any finite size $L$, it is possible to  start from $T$ and build 
in the disorder terms $\hat{\Delta}_l$ completely.  For that, it is 
convenient to work in the representation diagonalising $T$. 
Two ($i=1,2$ say) of the four labels on its eigenvalues $\lambda_i$ 
relate to the "hole" sector part $t_l$ of the transfer matrix.  The "particle" 
sector ($i=3,4$) couples to the other sector through the inhomogeneous 
term referred to above. The eigenvalues are conveniently parametrised as
$\lambda_{1,2} = \sqrt{h'/h}\, \exp (\pm iq)$, where $q$ satisfies 
$\Lambda = 1 - 2 \sqrt{h' h}\, \cos q$,  with a similar form (with 
$\Lambda$ changed sign) for $\lambda_3, \lambda_4$.

The formal expansion in powers of $\hat{\Delta}_l$ gives 
\begin{equation}
{\rm Tr}\; \prod_{l=1}^{L} T_l = \sum_{n=0}^{L} (A_n + B_n),,
\end{equation}
where $A_n$ and $B_n$ come respectively from the $i=1,2$, 
and $i=3,4$ terms  in the trace, and each involves sums over $n$
disorder variables.  For example,
\begin{equation} 
\label{multiple}
A_n = (4 \,\delta)^n\,  \left( h'/ h \right)^{L \over 2}
\underbrace{ \sum_{m_1 \ge 1}\,\sum_{m_2 \ge 1} ... 
\sum_{m_n \ge 1}}_{\sum_{i=1}^n\,m_i = L}\,
\zeta_{_{m_1}}\,\zeta_{_{m_1+m_2}}\,...\,
 \,\zeta_{_{m1+...+m_n}}\,
\prod_{i=1}^n \eta\,(m_i q)\,,
\end{equation}
where $\eta (mq)$ is a periodic function involving separations $m_i$ 
of the participating disorder sites and connected with competing 
aspects of the wavelength and scales set by the local disorder and bias.
$B_n$ involves a part similar to $A_n$ and a second more complex part 
generated by the inhomogeneous terms.

It is instructive to consider the approximation to the  multiple sum in 
Eq.\,(\ref{multiple}) coming from taking all $m_i$ as large.  
In this case the $q$-dependence in the functions  $\eta (m_i q)$ goes into
an overall exponential factor,  and the product of these factors comes
out of the multiple sum as a complex phase factor $\exp (\pm i\,L q)$ 
also accompanied by further $q$-independent factors. The summand is 
then just the product of disorder variables {$\zeta_{...}$} and the sum 
can be shown to be the coefficient of $x^n$ in the generating function 
$\prod_{l=1}^{L} (1 + x \zeta_l)$. Taken together with the other factors 
that leads to $\sum_{n=0}^{L} A_n$ being the exponential of an 
aggregated complex phase with real and imaginary parts 
$\sum_{l=1}^L \phi_l$ and $i\, L {\rm Re}\,(q)$, where
\begin{equation}
\label{phase}
\phi_l = \ln\, (h'_l/h_l)\,.
\end{equation}

For binary random rates the sum of $\phi_l$ is like a 
random walk. The case of most interest is when  the two possibilities 
have different signs (orientations) for the bias $h_l - h'_l$ , so steps occur 
in both directions. The same phase step (though aggregated in various 
different ways) occurs in the parts of the corresponding approximation 
for $\sum_{n=0}^{L} B_n$  coming from the inhomogeneous terms;
the remaining "homogeneous part" has negligible phase step.

However, in the approximation the location of states then becomes 
the same as for the pure system - coming just from $ i\,L {\rm Re}\,(q)$
since the disorder dependence of the phase is now confined to the real 
function $\phi_l $. So it will be necessary to generalise the result obtained 
by taking characteristic $m$'s as large, by retaining the registration 
between local disorder and aggregating phase which is provided by the 
functions $\eta (m_i q)$. This is most easily done by working with the 
fully-disordered transfer matrix (or directly from the original equations
of motion). That gives, as follows,  information on characteristic scales, 
domain structures, phase accumulations, and the region of validity of the
approximation just discussed. For the phases from $\sum_{n=0}^{L} 
A_n$ (the simplest illustration) only the $2 \times 2$ subpart $t_l$ 
corresponding to the hole sector of the  (fully-disordered) transfer matrix 
$T_l$ is needed.

The crucial aggregating phase variable is $\ln\, x_l$, where $x_l =
(a_{l-1}/a_l)$  is the ratio of complex amplitudes $a_l$ 
on successive sites. These $x_l$ ratios satisfy
\begin{equation}
\left(\begin{array}{c}
x_{l+1}^{-1}  \\
1
\end{array}\right) 
= t_l
\left(\begin{array}{c}
1  \\
x_{l}
\end{array}\right)\,.
\end{equation}
From this it follows that $x_l =g_l (h'_{l-1}/h_{l-1})$, where $g_l$ 
is such that $z_l \equiv (1 - g_l)^{-1}$ satisfies
\begin{equation}
z_{l+1} = [(h_l + \Lambda)z_l + h'_{l-1}]/[\Lambda z_l + h'_{l-1}]\,.
\end{equation}
From this (Mobius) map, and its first iteration yielding $z_{l+2}$ , 
it can readily be seen that when
\begin{equation}
\label{region}
 z^<_l   \equiv  \frac{h'_{l-1} }{h_l}\, [\,1+ {\cal O} (\Lambda)\,]
\ll \vert z_l \vert \ll \frac{h'_{l-1} }{\Lambda} \equiv z^>_l\,,
\end{equation}
(needing $\Lambda$ small) the variable $\chi_l \equiv \ln\, z_l$ "walks" 
according to $\chi_{l+1} \sim \chi_l + \ln\, (h_l/h'_{l-1})$ while the 
complex phase step in $a_l$ is $\ln\, x_l = \ln\, [g_l (h'_{l-1}/h_{l-1})] 
\sim \ln\, (h'_{l-1}/h_{l-1}) = \phi_{l-1}$,  as was found 
in Eq.\,(\ref{phase}) using 
the approximation on $\sum_{n=0}^{L} A_n$.  Near $z_l = z^>$ (or  
$z^<$), the mapped $z$ goes to $\infty$ (or 0, then 1,  giving a  
vanishing amplitude) and changes sign. This procedure thus provides the
region (\ref{region}) of validity of the approximation on $\sum_{n=0}^{L} 
A_n$, and generalises it: the behaviour at $ z = z^< ,  z^>$ is the proper 
consequence of the functions $\eta (m_i q)$, replacing the sign changes 
given by $\exp (\pm i\,L q)$ in the approximate result.

In the corresponding discussion for the structure and phases in 
$\sum_{n=0}^{L} B_n$ (coming from "particle" states, with complex 
amplitudes $a^+_l$), linearity allows a separation into two parts: 
(i) without, and (ii) with, the mixing to the states in the "hole" subspace 
just treated. The discussion for the first part ((i)) is similar to that 
just given, and leads to a walk step $\ln\, (a^+_{l-1}/a^+_l) \equiv
 \ln\, c^+_l$ with $c^+_l$ close to 1 if $y_l \equiv (1 - c^+_l)^{-1}$
satisfies a condition like (\ref{region}), with the same limits  $z^< ,\,  
z^>$ to leading order in $\Lambda$. For small $\Lambda$, $\varphi_l 
\equiv \ln\, y_l$  walks with the same step as $\chi_l = \ln\, z_l$, and has
 the same behaviour at the limiting values. For part (ii), the 
inhomogeneous term coming from the mixing to the states in the other 
subspace leads to the occurrence again of $z_l$ and  $x_l =(a_{l-1}/a_l)$,
in terms of which,  for $z^< \ll z_l \ll z^>$ (so $\Lambda$ is small and 
$x_l \sim (h'_{l-1}/ h_{l-1}) $) , the amplitude ratio 
is  $a^+_l/a^+_0 = 1 + \sum_{j=0}^{l-1} (\Gamma_l / h_l)$ where
\begin{equation}
\Gamma_l \equiv h_l ( a^+_{l+1} - a^+_l) = a_0 \prod_{j=0}^{l+1}
 \gamma_l^{-1} + A_0  \prod_{j=0}^l \gamma_l  , \; l \geq 1\,,
\end{equation}
with $\gamma_l = h'_l/ h_l$,  and $A_0 = \Gamma_0 - a_0/x_1$.  

The combination of these results from (i),(ii) gives the essence of the 
complicated result for the approximated $\sum_{n=0}^{L} B_n$, 
together with its validity criterion, and again generalises it as necessary
to obtain and quantize the imaginary part of the phase, to enumerate the
states:  the limiting boundary phase conditions provide the sign reversals
in amplitude and correspond to (reflecting and absorbing) boundary
conditions on the walks described above.

Since the walk steps of $\chi_l$ and $\varphi_l$ contain disorder 
we need the probability distributions  $P(\chi_l)$ and $P(\varphi_l)$, 
which are the same for small $\Lambda$.  The distributions depend on 
the boundary conditions, and are most easily treated in a continuum
(diffusion) picture. There, $l$ plays the role of time,  and the accumulating
$\chi_l$ or $\varphi_l$ plays the role of spatial coordinate.  From the walk 
steps described above,  within the limits set by Eq.\,(\ref{region}),  the
distributions satisfy a standard biased 
diffusion equation  having diffusion constant $D$ and bias $b$ given by 
\begin{eqnarray} 
\nonumber
D &=& \frac{1}{2}\, {\rm var\; \ln\,}\,(h_l / h'_l) \,\,,\\
\label{bias}
b &=& \langle \, \ln\,\,h_l\, \rangle - \langle \,\ln\,\,h'_l\, \rangle\,.
\end{eqnarray}

The boundary conditions take the biased problem to first-passage 
form [\onlinecite{Redner}] in which many reflections of the walk occur
before absorption is reached, with probability exponentially small in the 
traversal "length". The boundary conditions  set the limits to the 
excursion (of the walk), making the traversal "length" 
\begin{equation}
\ln\, z^> - \ln\, z^< = \ln\, (\Lambda_0 / \Lambda ) , 
\; \Lambda_0 \sim h\,,
\end{equation}
(long, as required for the continuum picture, in the critical regime of small
$\Lambda$). Consequently, the first passage "time" scale ${\cal L}
\sim L$, is proportional to $\exp [(|b|/D) \ln\, (\Lambda_0 /
 \Lambda )]$ , which can be written in  the scaling form 
$\Lambda^{-1/z}$ with dynamic exponent
\begin{equation}
\label{zeta}
z = D\,/ \vert \,b \,\vert.
\end{equation}
In the opposite case of large bias, applying for nearly homogeneous systems,
the effects of $({\rm Re}\; \Lambda)^{1/2}$  dominate making $z$ close to 2.

From the exponential distribution $P (\chi) \propto \exp[-\,(\,
\vert\, b\,\vert / D)\, \chi]$ for $\chi$ (and similarly $\varphi$), the 
energy distribution is appreciable only at $\chi = \ln\, z^<(\Lambda)$. 
That gives the exponential energy distribution
\begin{equation}
P(\Lambda) \propto \exp [-\, \Lambda \,\vert \,b\,\vert \,h'/ (D\,h)\,]\,,
\;\;{\rm for} \;\; \Lambda \gtrsim{\cal L}^{-z}\,.
\end{equation}

It can be seen that the dynamic exponent $z$ is non-universal because 
the distribution of the random rates allows both signs of the walk steps 
and is (in general) parameter-dependent. In particular, $z$ can diverge 
if the bias $b$ becomes zero (for example, at a critical concentration $p_c$ 
in the binary case studied below in Sec. IV). Then a stretched exponential 
scaling form emerges as follows. For the unbiased case the first passage 
problem has standard $\chi^2 \propto D \cal L $ relation between the "length"
and "time" scales. That makes $[\ln\,(\Lambda_0 / \Lambda)]^2 
\propto D \cal L$, consequently
\begin{equation}
\label{stretched}
\Lambda / \Lambda_0 \sim \exp \left(\,-c\, {\cal L}^{\beta}\right)\,,
\end{equation}
with $\beta = 1/2$ and $c = c_0 D^{1/2}$, with $c_0$ a numerical 
constant of order one.  In this case (where the bias vanishes) the Gaussian 
distribution for  the aggregated phase produces a gap distribution with 
exponential tails.  Consequently, for $b \rightarrow 0$ , the ratio of rms
gap to mean gap diverges.

\section {Numerical results}

\begin{figure}[htbp]
\vskip -2cm
\centering
\includegraphics[width=0.43\textwidth]{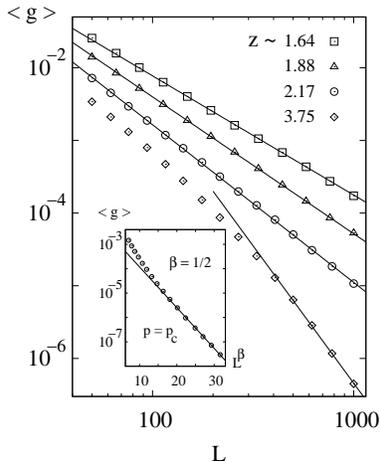}
\vskip -3.1cm
\caption{Average spectrum gap $\langle g \rangle$ for a binary distribution of
biases $\{0.6,\,-\,0.2\}$,  and instantaneous reactions ($R_{1,2} \equiv 1$). 
The concentration varies from top to bottom as $p =$ 0.5, 0.75, 0.3, 0.25.  
Dynamic exponents $z$ are read off from the slopes of the fitting lines. 
For displaying purposes, data have been shifted slightly upwards with 
respect to  $p = 0.25$. The inset shows the stretched exponential 
scaling expected  at $p_c \sim 0.2263$ under free boundaries.}
\label{FSS}
\end{figure}

To check out the above expectations, 
we now investigate numerically the finite size behavior of the gap of the
original evolution operator (\ref{Hamiltonian}) when averaged over 
independent disorder realizations. To this end, we diagonalize the 
$M_{\pm}$ matrices referred to in Eq.\,(\ref{matrices}) given in Sec. II,
and whose dimensions just grow linearly with the system size. 
For concreteness and  comparison with the analytical results of Sec. III, let us 
consider a binary distribution of bonds with parameters 
$B_{1,2} = \{R_{1,2}\,, \,h_{1,2}\}$, 
and probability  $P (B) = p \, \delta_{B, B_1} + (1-p)\,\delta_{B,B_2}$. For a 
given concentration $p$, we averaged spectra typically over 6000 samples 
of chains with lengths $50 \le L \le 1000$.  
For most $p$ the resulting  average gaps scale as $\langle g \rangle
\propto L^{-z}$ irrespective of parity, and eventually with  
{\it non-universal} dynamic exponents $z = z (p)$, as it was established by 
Eq.\,(\ref{zeta}) in Sec. III. This situation is exhibited in Fig.\,\ref{FSS},  
where we show a characteristic case of instantaneous reactions under 
orientation changes in diffusion biases $h-h'$, these latter satisfying
the free fermion constraint (\ref{constraint}) imposed throughout.

On approaching the critical regime defined by the unbiased diffusion
equation referred to earlier on in Sec. III, finite size effects become 
progressively pronounced until at a critical concentration $p_c$, given by
\begin{equation}
\label{critical}
p_c = \left[\,1 - \frac{\ln \left(\frac{1-h_1}{h1}\right)}
{\ln \left(\frac{1-h_2}{h2}\right)}\,\right]^{-1}, \;\;
h_1\, ^<_> \; \frac{1}{2}\; ^<_>\,h_2\,,
\end{equation}
[\,i.e. $b = 0$, in  Eq.\,(\ref{bias})\,], the length scaling of the gap crosses over
to the stretched exponential form conjectured above in Eq.\,(\ref{stretched}). 
This is corroborated in the inset of Fig.\,\ref{FSS} which
\begin{figure}[htbp]
\vskip -2.3cm
\centering
\includegraphics[height=0.45\textheight]{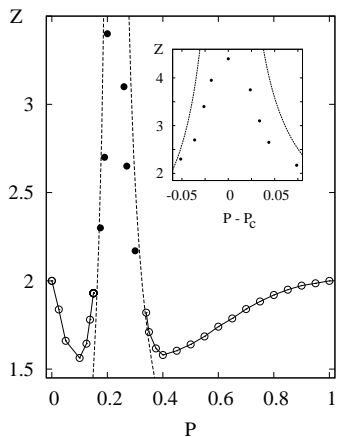} 
\vskip -2.7cm
\caption{Non-universal dynamic exponents of the binary distribution
of Fig.\,\ref{FSS}. Filled circles near $p_c$ stand for lower bounds of $z$. 
In the inset these latter support the divergence trend  of the analytical
results close to criticality [Eq.\,(\ref{zeta}) in dashed curves\,].}
\label{non-universal} 
\end{figure}
for large sizes (and open boundaries [\onlinecite{PBC}] ), supports the 
universal stretching factor $\beta = 1/2$ involved in (\ref{stretched}). 
Thus, in nearing $p_c$ the stochastic dynamics slows down dramatically 
which, in line with the approach of the preceding section, is reflected in the 
abrupt increase of dynamic exponents  observed in Fig.\,\ref{non-universal}.
As we have seen in Eq.\,(\ref{zeta}), for $p \to p_c$ these latter can be 
arbitrarily large in the thermodynamic limit, but in practice they are 
ultimately cutoff by our available system sizes. In this regard, notice that the 
trend of slopes obtained near the critical region (e.g. lowermost curve of 
Fig.\,\ref{FSS}), constitutes a numerical lower bound for these exponents 
(filled symbols of Fig.\,\ref{non-universal}).
\begin{figure}[htbp]
\vskip -2.5cm
\centering
\includegraphics[width=0.47\textwidth]{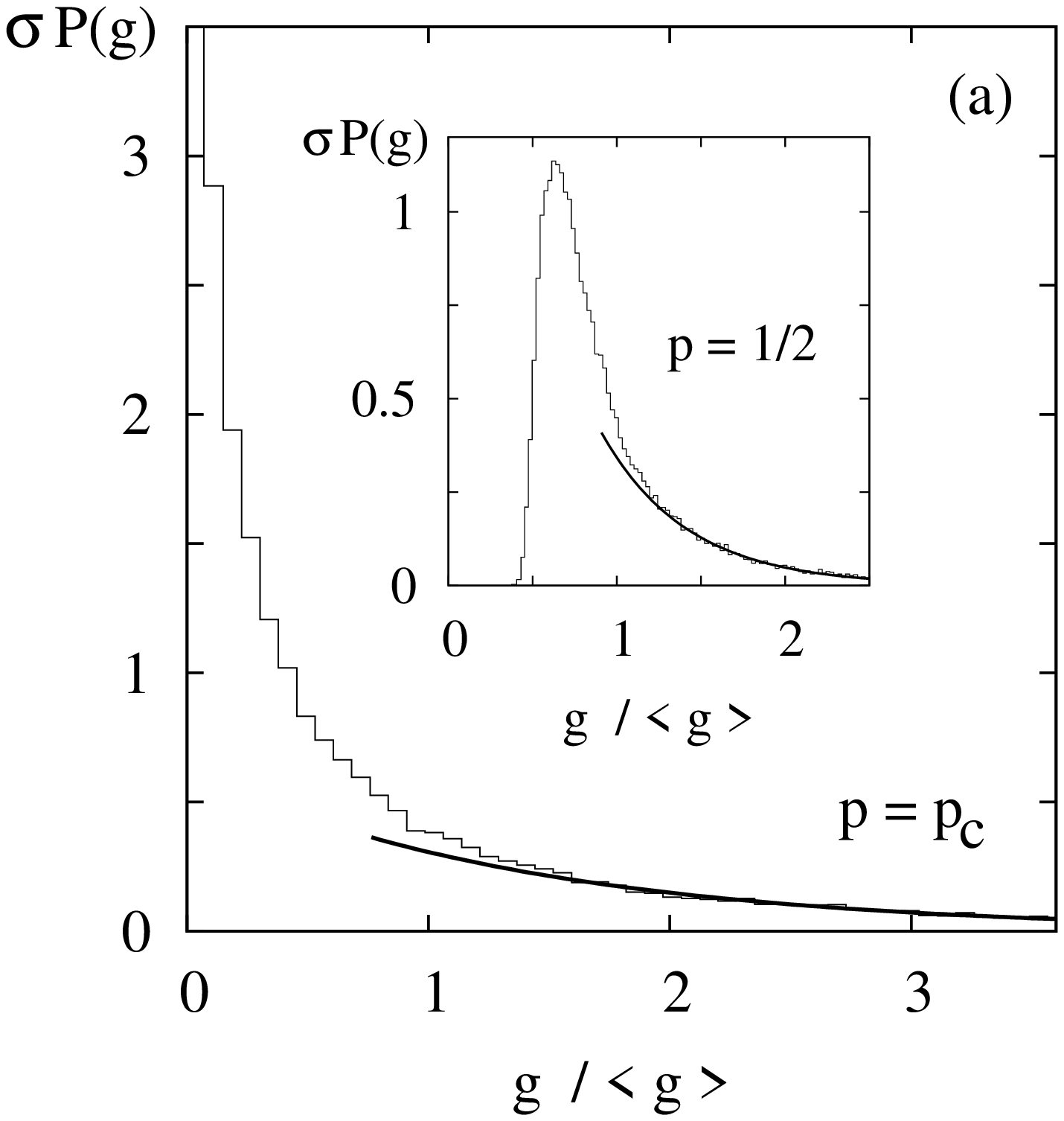}
\vskip -5.5cm 
\centering
\includegraphics[width=0.47\textwidth]{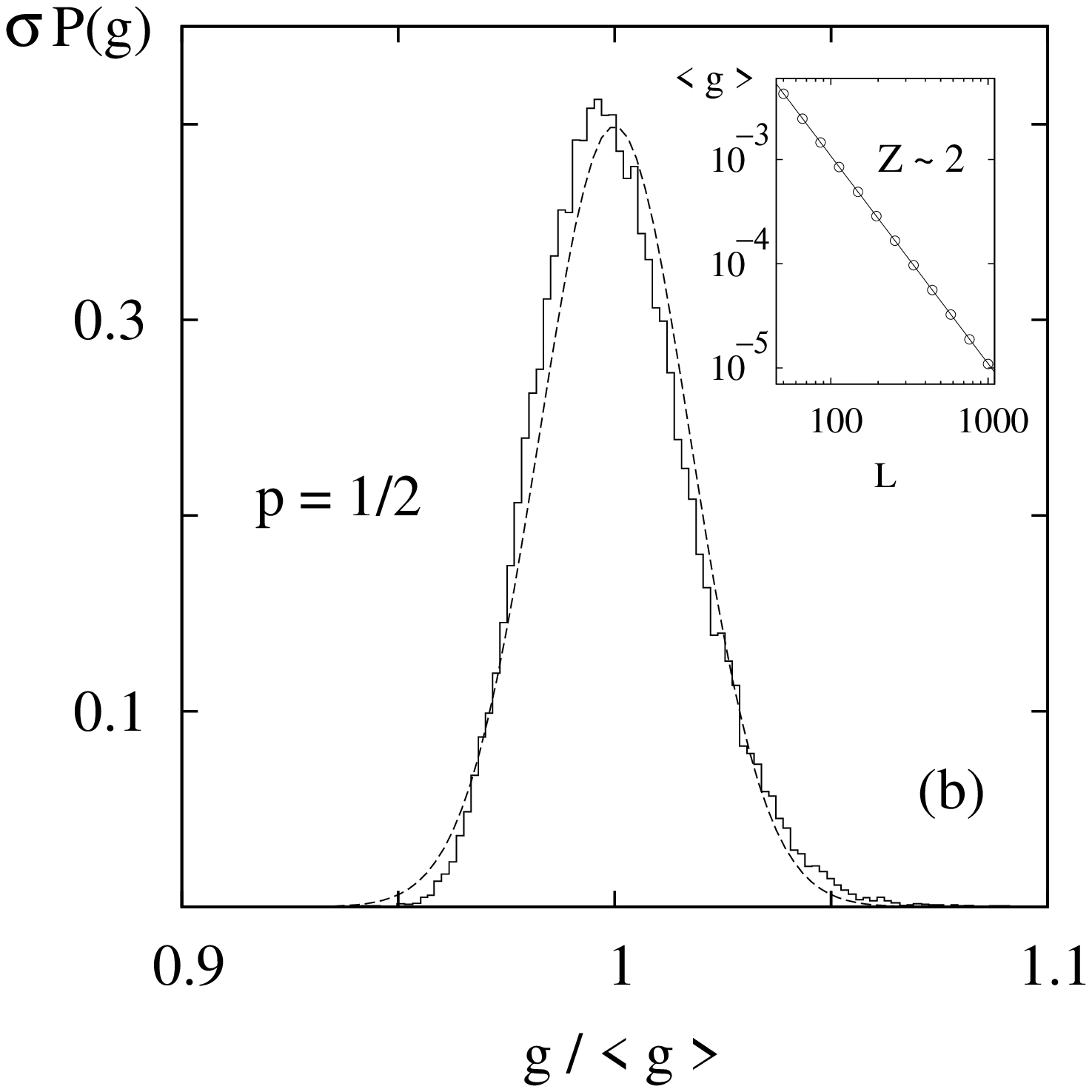}
\vskip -3.5cm
\caption{Distributions of gaps $(g)$ for binary disorder over 500 sites.
In (a) the parameter rates are taken  as in Fig.\,\ref{FSS}; the main 
panel shows (part of) the case $p = p_c$ ($\sigma / \langle g \rangle
\sim 2$), whereas the inset displays a characteristic non-critical 
situation ($p = 0.5,\, \sigma/\langle g \rangle \sim 0.6$).
Both results have exponential tails (in bold lines). (b) shows a typical 
well defined bias case $(\{0.6,0.2\},\, p=0.5)$, exhibiting instead a
narrow distribution closely following a Gaussian (in dashed line, 
$\sigma / \langle g \rangle \sim 0.02$).  As is shown in the inset, 
here dynamic  exponents are diffusive.}
\label{DOG}
\end{figure}

Also, in agreement with the theoretical analysis, it is seen that the distribution
of spectral gaps becomes broadest at $p_c$ (maximum variance $\sigma$). 
This is illustrated in Fig.\,\ref{DOG}a, after binning gaps in 800 intervals  
over 60.000 disorder realizations. By contrast, chains with non-fluctuating 
bias orientations have rather narrow gap distributions, and typically diffusive  
dynamic exponents ($z =  2)$, regardless of the value of $p$.  
This characteristic scenario already anticipated in Sec. III,  is exhibited in  
Fig.\,\ref{DOG}b.

\begin{figure}[htbp]
\vskip -2cm
\centering
\includegraphics[width=0.43\textwidth]{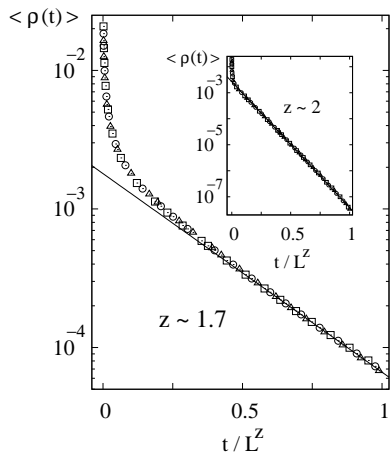}
\vskip -2.7cm
\caption{Density decay for the distribution of Fig.\,\ref{FSS} exhibiting a 
non-universal dynamic exponent at $p = 0.5$. The inset shows the case
of a well defined sign of binary biases $\{0.6,\,0.2\}$ (as in 
Fig.\,\ref{DOG}b), displaying  instead a typical diffusive exponent. 
Squares denote data for $L = 800$, circles 400, and triangles 200. 
To compare slopes, the data of these latter two sizes were shifted 
downwards.}
\label{MCarlo}
\end{figure}

{\it Simulations.}-- 
To allow for an independent examination of nonuniversal aspects, we also 
carried out numerical simulations in periodic chains of  200, 400 and 800 
sites (without duplication). Averages of particle densities were taken over 
$10^3$ histories of  $10^3$ samples with binary disorder. Starting from 
homogeneous densities $\rho$, our results in Fig.\,\ref{MCarlo} clearly 
indicate an asymptotic decay of the form $\langle \rho_L (t) \,\rangle 
\propto e^{-t / \tau_L}$ with a relaxation time $\tau_L \propto L^z$. 
As expected, the slope collapse follows closely the non-universal dynamic 
exponent already identified by direct diagonalization ($z \sim 1.64$, 
in Fig.\,\ref{FSS}). In turn, the best scaling behavior for well oriented 
biases is obtained with standard diffusive exponents, which is also in 
agreement with our previous numerical results (inset of Fig.\,\ref{DOG}b).

\begin{figure}[htbp]
\vskip -0.25cm
\centering
\includegraphics[width=0.27\textwidth]{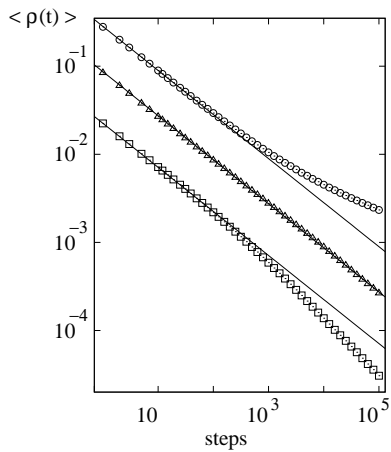}
\vskip 0.5cm
\caption{Density decay for the binary distribution considered above, using 
$L = 10^4$ and different concentrations. From top to bottom these vary as 
$p_c \sim 0.2263$ (circles), $0.15$ (triangles) , and $0.5$ (squares). For 
comparisons of large times trends, the data of the latter two concentrations
were shifted downwards. Solid lines are fitted with slopes -1/2.}
\label{decay}
\end{figure}

Finally, it is interesting to comment on the regime $t \ll L^z$. This
latter is addressed in Fig.\,\ref{decay} where the density decay below,
at, and above $p_c$ is exhibited for much larger systems. After averaging 
20 samples over 100 histories each, it is observed that in all cases there is
an early stage showing the typical $t^{-1/2}$ decay. However, at $p = p_c$
($p > p_c$), an incipient crossover to a slower (faster) decay clearly emerges
(cf. [\onlinecite{Gunter,Krapivsky}]).  It would take further evolution decades
to estimate the actual form of such regimes, although their nonuniversal 
characters are already in line with our findings for $z(p)$. The situation for 
$p < p_c$ remains unclear as the existence of a very slow crossover can not 
be ruled out.

\section{Conclusions}

To summarise, after identifying phase steps, and limiting phases and their 
relationship to generalised biased random walks, in Sec. III we have outlined
how the character (power law or stretched exponential) of the non-universal
critical dynamics and in particular the critical exponents and their 
non-universality are determined by the associated diffusion constant and bias 
(in the case where the fundamental phase step can have either sign). 

Among the specific results given above are: in the general biased case the
dynamic exponent $z$ is $D/\vert b \vert$ and associated distributions of the
energy gap are exponential. For vanishing bias $z$ diverges and the usual 
critical power law behaviour is replaced by a stretched exponential form,
possibly signaling the emergence of a glassy dynamic, while the ratio of rms 
gap to mean gap diverges. All those findings were corroborated numerically 
using the doubling fermion approach given in Sec. II and Monte Carlo 
simulations of Sec. IV.

Concerning occurrence of both positive and negative phase steps, that is 
connected, in a given configuration of randomness, to a succession of 
domains of alternating bias, whose boundary sites alternate between 
being traps or repellers for particles. That causes some particles to 
wander ineffectively for long times before finding a partner for pair 
annihilation (see  Fig.\,\ref{snap}), leading to large or divergent dynamic 
exponents (see Figs.\,\ref{FSS} and \ref{non-universal}).

Among many important things not covered are: 
the full parameter-dependence of dynamic exponent $z$ from pure value 
to divergence where the bias vanishes (e.g. in the case of the binary 
distribution of rates, with concentration parameter p, where $z(p)$ is 
not monotonic for $p$ in $[0\,,\,p_c]$ or in $[p_c\,,\,1]$);  the effect of
open boundaries versus PBC (can be very significant in driven non-equilibrium
systems) - particularly here regarding the stretching exponent for zero bias
 - needs further investigation.

\section*{Acknowledgments}

Numerical evaluations were carried out on the HPC Bose Cluster
at IFLP.\,  MDG and GLR acknowledge discussions with C.A. Lamas and
H.D. Rosales, and support of CONICET, Argentina, under grants PIP 5037
and PICT ANCYPT 20350. RBS acknowledges discussions with J.T. Chalker, 
and EPSRC support under the Oxford grant EP/D050952/1.



\begin{thebibliography}{99}
\bibitem{PBC} Under PBC however, our diagonalizations
suggest instead a stretching factor $\beta \sim 0.3$.
\end{thebibliography}
\end{document}